# Evolution of Magnetic and Superconducting Fluctuations with Doping of High-$T_c$ Superconductors

G. Blumberg,* Moonsoo Kang, M. V. Klein, K. Kadowaki, C. Kendziora

Electronic Raman scattering from high- and low-energy excitations was studied as a function of temperature, extent of hole doping, and energy of the incident photons in $Bi_2Sr_2CaCu_2O_{8\pm\delta}$ superconductors. For underdoped superconductors, short-range antiferromagnetic (AF) correlations were found to persist with hole doping, and doped single holes were found to be incoherent in the AF environment. Above the superconducting (SC) transition temperature $T_c$, the system exhibited a sharp Raman resonance of $B_{1g}$ symmetry and energy of 75 millielectron-volts and a pseudogap for electron-hole excitations below 75 millielectron-volts, a manifestation of a partially coherent state forming from doped incoherent quasi particles. The occupancy of the coherent state increases with cooling until phase ordering at $T_c$ produces a global SC state.

The normal state properties of doped cuprate high-temperature superconductors are markedly different from those of conventional metals and are usually viewed as manifestations of strong electron-electron correlations. These correlations cause the AF state in undoped cuprates. In the SC state, inelastic neutron scattering reveals a novel resonant magnetic excitation (*1, 2*). Here, we report a Raman active resonance that is present in doped cuprate superconductors above $T_c$ and gains strength with cooling.

The doping phase diagram of high-$T_c$ cuprate superconductors and parent materials can be divided into four doping regions: (i) a spin $S = ½$ AF insulator with the spin localized on Cu atoms of the $CuO_2$ planes, a strong superexchange constant $J \approx 125$ meV, and a high Néel temperature $T_N \approx 300$ K that rapidly drops with hole doping in the $CuO_2$ planes until a metal-insulator transition is reached; (ii) an underdoped superconductor, where $T_c$ increases with increasing hole doping; (iii) an optimally doped superconductor, where $T_c$ reaches its maximum; and (iv) an overdoped superconductor, where the cuprate becomes a better metal as hole doping progresses and $T_c$ decreases. Two phase transition lines separate the "normal" high-temperature state from the low-temperature AF and SC phases.

In addition to these AF and SC phase transitions, the phase diagram contains two crossover lines that start at high temperatures for low doping, decrease in temperature for the underdoped region, and merge at the SC transition line for the slightly overdoped SC (*3*). At the upper crossover temperature, short-range AF correlations develop, which, for very low doping and lower temperatures, evolve into the long-range ordered AF state. The second, lower, crossover temperature is experimentally suggested for the underdoped materials. At this temperature, the system develops a suppression of the density of low-energy excited states, characterized as the result of a "pseudogap" (*4*). The gap removes only a fraction of the states at the Fermi energy ($E_F$). The material remains metallic and, moreover, shows an increase in the component of the electrical conductivity parallel to the $CuO_2$ planes (*5*). Nuclear magnetic resonance measurements revealed a drop in the Knight shift as the temperature $T$ is decreased above $T_c$ (*6*), which suggests that in these materials the pairing of electronic spins into singlets takes place at temperatures higher than $T_c$. The copper nuclear spin relaxation rate divided by temperature first rises to a broad maximum for decreasing $T > T_c$, and then falls (*6*), suggesting that a pseudogap opens up in the low-energy electronic excitation spectrum. Suppression of the spin susceptibility above $T_c$ has been observed in inelastic neutron scattering experiments (*1*). A downturn above $T_c$ in the electronic density of states (DOS) at $E_F$ has been suggested from high-resolution heat capacity measurements (*7*). Recent optical studies of underdoped cuprates reveal a drop in the low-frequency scattering rate, and thus an increase in the coherence of the electronic system, in the pseudogap regime (*8*). Results from angle-resolved photoemission spectroscopy (ARPES) (*9–11*) and tunneling (*12*) confirm the normal state gap-like depression of the electronic DOS of underdoped cuprates. The momentum dependence of this pseudogap resembles that of the $d_{x^2-y^2}$ gap observed in the SC state (*9, 10*).

What causes the pseudogap and the unconventional normal state properties of underdoped cuprates? What is the nature of the SC pairing state? The AF and SC properties of cuprates are strongly related (*13*). The SC coherence length $\xi_{SC}$ is on the order of a few lattice spacings, considerably shorter than in conventional superconductors. The persistence of short-range AF correlations with doping, $\xi_{AF} \gtrsim \xi_{SC}$, has been thought to lead to an effective pairing mechanism (*14, 15*). It is also thought that AF correlations that do not vanish with doping are responsible for both the unconventional normal state properties and the high-$T_c$ superconductivity phenomenon (*16–20*).

Underdoped SC cuprates have low carrier density. Because of the persistence of short-range AF spin correlations, the doped holes are very heavy (*21*) or quasilocalized (*15*). The mean free path of the holes is shorter than their de Broglie wavelength, which brings the underdoped cuprates into a class of "bad metals" (*22*). London penetration depth and optical conductivity measurements (*23*) show that the Drude weight in the normal state as well as the superfluid density in the SC state are proportional to the carrier doping. Moreover, $T_c$ and the superfluid density are linearly related (*24*). For cuprates, a rough estimate of parameter $k_F\xi_{SC}$ (where $k_F$ is the Fermi wave vector) leads to values of about 3 to 20, two orders of magnitude smaller than for conventional Bardeen-Cooper-Schrieffer (BCS) superconductors. The SC transition for cuprates occurs around the temperature at which the thermal de Broglie wavelength of the pairs is two to six times the average interpair separation (*24*). A pair size comparable to the average interparticle spacing $k_F^{-1}$ brings cuprates to an intermediate regime between the BCS limit of large overlapping Cooper pairs ($k_F\xi_{SC} \gg 1$) and that of Bose-Einstein (BE) condensation of composite bosons ($k_F\xi_{SC} \ll 1$) consisting of tightly bound fermion pairs (*24, 25*). As a consequence of the low carrier density and the short SC correlation length, the transition to the SC state may not display typical BCS behavior. In underdoped cuprates, $T_c$ may be strongly suppressed with respect to the pairing temperature and may be determined by phase fluctuations (*22, 26, 27*).

Electronic Raman scattering is a local high-energy probe for the short-range AF correlations as well as for the SC order

G. Blumberg, NSF Science and Technology Center for Superconductivity and Department of Physics, University of Illinois at Urbana-Champaign, Urbana, IL 61801, USA, and Institute of Chemical Physics and Biophysics, Tallinn EE0001, Estonia.
M. Kang and M. V. Klein, NSF Science and Technology Center for Superconductivity and Department of Physics, University of Illinois at Urbana-Champaign, Urbana, IL 61801, USA.
K. Kadowaki, Institute of Materials Science, University of Tsukuba, Tsukuba, Ibaraki 305, Japan.
C. Kendziora, Naval Research Laboratory (Code 6653), Washington, DC 20375, USA.

*To whom correspondence should be addressed (at University of Illinois). E-mail: blumberg@uiuc.edu

parameter in doped SC samples through excitation across the SC gap. We studied both effects in $Bi_2Sr_2CaCu_2O_{8\pm\delta}$ crystals and found that for underdoped samples, the short-range AF correlations persist with hole doping. Furthermore, we interpreted the presence of a Raman peak of $B_{1g}$ symmetry at ~75 meV as evidence that incoherent doped holes bind above $T_c$ in a long-lived collective state with a sharp resonance of $B_{1g}$ ($d_{x^2-y^2}$) symmetry and a binding energy of 75 meV. The temperature dependence of this peak shows that this state gains phase coherence at $T_c$ and participates in the collective SC state. The binding of incoherent quasi particles (QPs) in the coherent state reduces the low-frequency scattering rate and leads to a pseudogap in the spectra.

Two-magnon (2-M) Raman scattering directly probes short-wavelength magnetic fluctuations, which may exist without long-range AF order (28, 29). The Raman process takes place through a photon-stimulated virtual charge transfer (CT) excitation that exchanges two Cu spins. This process may also be described as the creation of two interacting magnons. The CT excitation is the same one that virtually produces the spin superexchange constant $J$. In the AF environment with a correlation length $\xi_{AF}$ covering two to three lattice constants, the spin exchange process requires an energy of ~3$J$. The magnetic Raman scattering peak position, intensity, and shape provide information about fluctuations in a state of short-range AF order (29).

Electronic Raman scattering by charge fluctuations in metals arises from electron-hole (e-h) excitations near the Fermi surface. For a normal Fermi liquid model of the cuprates, the scattering would have finite intensity only at very low frequencies. For strongly correlated systems, incoherent QP scattering leads to finite Raman intensity over a broad frequency region (30, 31), and the intensity can be used as a measure of the incoherent scattering. Indeed, for cuprates in the normal state, an almost frequency-independent Raman continuum has been observed that extends to at least 2 eV. The continuum strength is dependent on excitation wavelength and weakens with doping (29, 32, 33). In the SC state of optimally doped and overdoped cuprates, the low-frequency tail of the Raman continuum changes to reflect the SC effects. The opening of a SC gap reduces the incoherent inelastic scattering processes and reduces the strength of the continuum. Freed from the heavy damping, the QPs now show a gap in their spectral function. Thus, the electronic Raman spectrum of e-h pair excitations acquires the so-called 2$\Delta$ peak as a result of excitations across anisotropic gap 2$\Delta$(**k**), where **k** is a wave vector on the Fermi surface. A recent resonance Raman study concluded that the 2$\Delta$ peak is caused by renormalization of the continuum in the SC state (34). For underdoped cuprates, only a relatively weak peak has been observed within the strong Raman continuum (35); its energy does not scale with sample $T_c$, and its origin has been unclear.

Bi-2212 single crystals were grown and postannealed as described (9, 36). The underdoped samples with $T_c$ = 83 K and $T_c$ < 10 K (showing a transition onset through magnetization below 10 K) were the same single crystals as in (9). Raman measurements were made with systems described in (33, 34). The presented data were taken in $xy$ scattering geometry, giving for Bi-2212 mainly Raman spectra of $B_{1g}$ symmetry. For magnetic excitations, the $B_{1g}$ scattering channel couples to pairs of short-wavelength magnons near the magnetic Brillouin zone boundary (29). For the e-h excitations near the Fermi surface, the Raman form factor for $B_{1g}$ symmetry is peaked for **k** near the antinode wave vectors $\{{\bf k}_{an}\}$ = [(0, $\pm\pi/a$) and ($\pm\pi/a$, 0)] (37), where the anisotropic SC gap magnitude is believed to reach its maximum value $\Delta_{max}$. Indeed, the $B_{1g}$ scattering geometry reveals interesting excitations in both magnetic and e-h excitation channels.

The high-energy part of $B_{1g}$ electronic Raman scattering spectra at room temperature is shown in Fig. 1 as a function of hole doping and excitation energy. The spectrum from the AF insulator (Y-doped Bi-2212 crystal) exhibits a band peaked at ~2860 cm$^{-1}$ and is assigned to scattering by 2-Ms, that is, the photon-induced superexchange of two spins on the two nearest-neighbor Cu $3d^9$-orbital sites (through the intervening 2$p$ oxygen orbital). The probability of the virtual intermediate CT process is expected to be resonantly enhanced when the incoming photon energy $\hbar\omega_L$ approaches the Cu 3$d$ – O 2$p$ CT energy $E_{CT}$ (38). As a result of the superexchange, each of two exchanged spins sees three ferromagnetically aligned neighbors, at a cost of ~3$J$ owing to the Heisenberg interaction energy $J\Sigma_{\langle i,j\rangle}({\bf S}_i\cdot{\bf S}_j-\frac{1}{4})$, where ${\bf S}_i$ is the spin on site $i$ and the summation is over near-neighbor Cu pairs. Thus, for the AF insulators, the 2-M peak position yields an estimate of $J \approx$ 125 meV. The probability of the superexchange process (the peak intensity) may be tuned by the incoming photon energy, showing enhancement for incoming photon energies close to the CT energy. Dispersion of the e-h band and of the magnon excitations puts an additional constraint on the most resonant superexchange process, which becomes most probable for those incoming photon energies that approach the upper edge of the e-h excitation spectrum (39). Resonance enhancement of the 2-M scattering yields information about $E_{CT}$ and the e-h band dispersion (29).

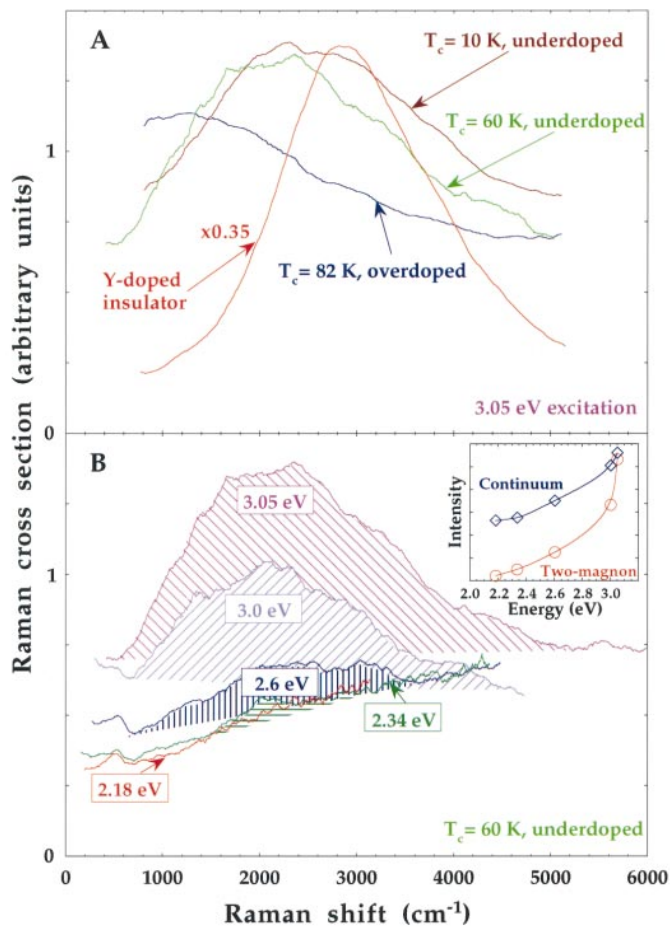

**Fig. 1.** The $B_{1g}$ continuum and two-magnon Raman scattering spectra at room temperature for $Bi_2Sr_2CaCu_2O_{8\pm\delta}$ as a function of (**A**) doping for 3.05 eV excitation and (**B**) excitation for underdoped $T_c$ = 60 K superconductor. The two-magnon intensity is shown by shading. A slope in the continuum is caused by the contribution of luminescence in the near infrared. Inset: the intensity of the two-magnon peak (circles) and of the continuum under it (diamonds) as a function of excitation energy.

For doped superconductors, the spectra exhibit a background continuum plus a broad peak, which, similar to the AF case, has been assigned to the double spin-flip excitation in the short-range AF environment. For the process of two-spin superexchange to require the full $3J$ energy cost, spins on six further Cu neighbor sites must show AF alignment. The doped holes are believed to form singlets with the holes that would otherwise form local AF order (40). This screens the effective spin moment on the doped Cu site and leads to a reduction of the spin superexchange energy in the vicinity of holes. With doping (Fig. 1A), the 2-M scattering peak broadens, weakens, and shifts to lower frequency. The existence of the peak in the SC cuprates indicates the persistence of a local short-range AF order extending for a few lattice spacings.

The continuum and the 2-M intensity depend on the incoming photon energy (inset, Fig. 1B), which confirms the resonance Raman scattering regime for the visible excitation range. Similar to the behavior of insulating cuprates (29), the 2-M intensity strongly resonates around $\hbar\omega_L^{res} \approx$ 3.1 eV excitation, confirming that the bottom of the valence band and the top of the conduction band are not changed by doping. The resonance provides an estimate for the bandwidth of e-h excitations across the CT gap as $\hbar\omega_L^{res} - E_{CT} \approx 1.4$ eV [an optical study (41) gives an estimate for $E_{CT} \approx 1.72$ eV] and for effective copper nearest-neighbor-hopping matrix element $t \approx 320$ meV because $\hbar\omega_L^{res} \approx 2[(E_{CT}/2)^2 + 16t^2]^{1/2}$. The continuum intensity also exhibits resonance enhancement toward violet excitation, but in contrast to the magnetic peak, this intensity does not drop as sharply toward the red, and the continuum scattering efficiency differs by only a factor of ~2 for the opposite ends of the visible excitation range.

The low-energy $B_{1g}$ Raman scattering spectra are shown in Figs. 2 and 3 as a function of doping, temperature, and excitation energy (42). The most prominent feature of the spectra is the electronic continuum. For blue excitations, the continuum is superimposed on the $\mathbf{q} \approx 0$ Raman-allowed optical phonons. The phononic scattering resonates more strongly toward the violet excitation than does the electronic continuum. This enables us to study clean continuum spectra without interfering phononic (and magnetic) excitations by using lower energy (red) excitation.

According to the existing phenomenology for the continuum (30, 43) and allowing for the symmetry of the $B_{1g}$ Raman form factor, we believe the $B_{1g}$ Raman continuum intensity, $I(\omega,T)$, to be proportional to the incoherent QP inverse lifetime $\tau_{\{k_{an}\}}^{-1}(\omega,T) = \Gamma_{\{k_{an}\}}(\omega,T) \approx [(\alpha\omega)^2 + (\beta T)^2]^{1/2}$ for $\mathbf{k}$ in the vicinity of wave vectors $\{\mathbf{k}_{an}\}$: $I(\omega,T) \propto [1 + n(\omega,T)]\omega\Gamma/(\omega^2 + \Gamma^2)$, where $n(\omega,T)$ is the Bose factor and $\alpha$ and $\beta$ are phenomenological parameters of order unity. The continuum extends to a few electron volts, an energy scale comparable to the width of the e-h excitation spectra (Fig. 1). For higher temperatures, the continuum starts from very low frequencies (Figs. 2 and 3), affirming strong incoherent scattering even for low-lying e-h excitations. This observation is consistent with ARPES studies (9, 11) showing that for wave vectors in the vicinity of $\{\mathbf{k}_{an}\}$, underdoped materials at higher temperatures reveal ill-defined QP peaks on a strong flat background. The holes cannot propagate freely on the short-range AF background; the strong correlations enforce incoherent excitations over the whole QP energy band range (15, 21).

Cooling the underdoped samples gradually strengthens a remarkably sharp scattering peak at ~600 cm$^{-1}$ (75 meV). The integrated (above the continuum line) intensity of the peak contains just a few percent of the integrated 2-M scattering intensity. This peak has $B_{1g}$ symmetry and is not present for polarized $xx$-scattering geometry ($A_{1g} + B_{2g}$). The peak position shows little temperature or doping dependence. We believe the origin of the peak is electronic rather than phononic. There are no allowed $B_{1g}$ $\mathbf{q} \approx 0$ phonon modes near 600 cm$^{-1}$ in optimally doped Bi-2212 (44). The 600 cm$^{-1}$ peak does not show strong intensity reduction for red relative to violet excitation, as shown by the phononic peaks (compare Figs. 2A and 2B). Assignment of the peak to an overtone or combination phonon mode is very unlikely, as these two-phonon peaks are almost always stronger in $A_{1g}$ symmetry than in any other Raman symmetry. Raman-active impurity-induced phonon modes are also usually stronger in $A_{1g}$ symmetry. Phonon Raman scattering is governed by a Bose factor, which causes the intensity to decrease with decreasing temperatures, opposite to the behavior observed here (see below). However, the 600 cm$^{-1}$ peak appears to be related to the features seen in one-electron spectroscopies. It occurs at about twice the energy of the QP peak observed by ARPES at low temperatures for **k** near $\{k_{an}\}$ points (9). Tunneling data on the underdoped samples exhibit peaks in conductance curves at $\sim \pm 36$ meV (45).

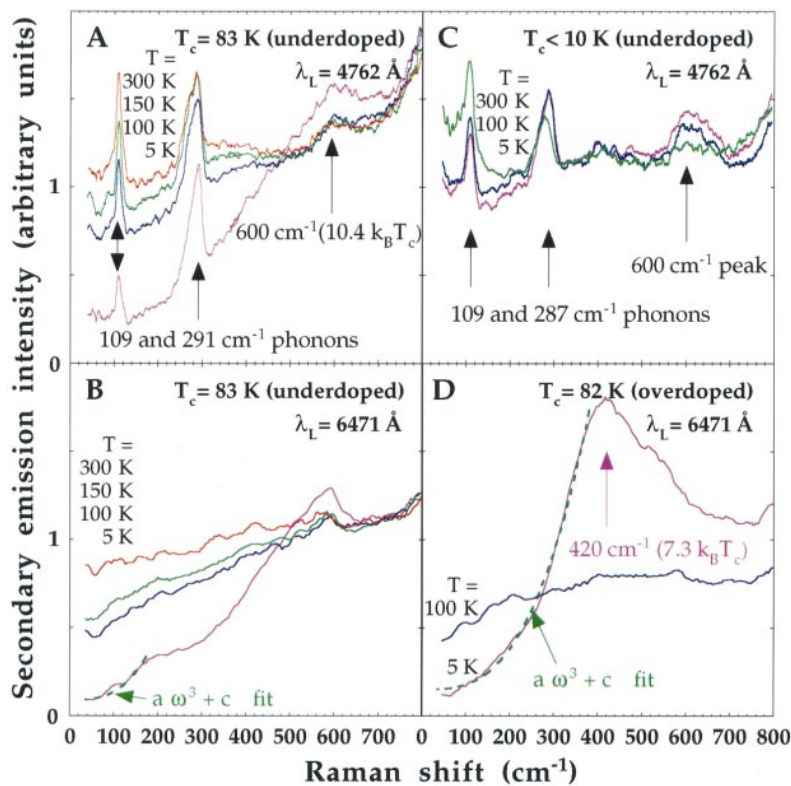

**Fig. 2.** The low-energy portion of $B_{1g}$ Raman scattering spectra for Bi$_2$Sr$_2$CaCu$_2$O$_{8\pm\delta}$ as a function of doping [(**A** and **B**) for underdoped sample with $T_c = 83$ K, (**C**) for underdoped sample with the SC transition onset at 10 K, and (**D**) for overdoped sample with $T_c = 82$ K], temperature, and excitation wavelength $\lambda_L$. The phononic modes resonate with blue excitation (A and C) and are not present in the spectra with red excitation (B and D). The 600 cm$^{-1}$ peak is already present in the room-temperature spectra of all underdoped samples and its intensity strengthens with cooling, whereas the electronic Raman scattering at low frequencies shows intensity reduction. Below $T_c$ for samples with higher dopings, the 600 cm$^{-1}$ peak develops in a 2Δ peak. At low temperatures the low-frequency tail of the 2Δ peak obeys cubic power law, shown by fit in (B) and (D).

The narrow width of the 600 cm$^{-1}$ peak in the Raman spectra ($\Gamma_{peak} \lesssim 50$ cm$^{-1}$) is more than an order of magnitude smaller than the inverse QP lifetime $\Gamma_{\{k_{an}\}}$ (600 cm$^{-1}$, T) associated with the continuum. The two very different lifetimes for the 600 cm$^{-1}$ mode and the continuum suggest that we are essentially dealing with a two-component system: a spectroscopically well-defined, long-lived 600 cm$^{-1}$ mode on top of the strong incoherent background.

The temperature dependence of the integrated intensity of the 600 cm$^{-1}$ mode (integrated above the continuum line) is shown in Fig. 4 for the three underdoped samples with $T_c$ = <10, 65, and 83 K (spectra shown in Figs. 2, B and C, and 3). The temperature dependence of the low-frequency continuum intensity is also shown. The integrated intensity of the 600 cm$^{-1}$ mode smoothly increases with cooling until the temperature reaches $T_c$, where the intensity shows a sudden enhancement. With cooling, the low-frequency portion of the continuum simultaneously shows intensity reduction. This covers a broader frequency range than can be explained by the temperature-dependent Bose factor. The intensity reduction is an indication of a drop in the low-frequency ($\omega < 600$ cm$^{-1}$) inverse lifetime $\tau_{\{k_{an}\}}^{-1}(\omega, T)$. The suppression of the low-frequency spectral weight in Raman spectra is similar to the pseudogap observed in spin-excitation spectra (1, 6) as well as in optical (8), ARPES (9–11), and tunneling (12) studies.

Various scenarios have been proposed to explain the pseudogap behavior. Within the nearly AF Fermi liquid model, Pines et al. suggested that the strong magnetic interaction, peaked near the commensurate AF wave vector $\mathbf{Q} \equiv (\pi/a, \pi/a)$, between the QPs in the vicinity of $\{k_{an}\}$ points leads to the formation of a precursor to a spin-density wave state with a pseudogap (17). Lee et al. interpreted the pseudogap as a spin excitation gap in a scenario where the electrons are decomposed into fermions that carry spin and bosons that represent charged vacancies. For the underdoped materials, the fermions become paired at some temperature above $T_c$, leading to the pseudogap (18). Emery et al. suggested the mechanism of pairing as a form of internal magnetic proximity effect in which a spin gap is generated in phase-separated AF regions through spatial confinement by charge stripes, then communicated to the stripes by pair hopping. The predicted collective excitations of the quasi–one-dimensional AF regions of the material include a neutron-scattering active charge-0, spin-1 magnon mode with an energy gap of the order of the SC gap plus a Raman active charge-0, spin-0 mode with $\sim \sqrt{3}$ higher energy (19). Uemura et al. (24) and Randeria et al. (25) proposed a preformed pairs scenario of a superconductor with strongly interacting dilute carriers. It has been suggested that in the presence of the short-range AF background $\xi_{AF} \gtrsim \xi_{SC}$, pairs of doped holes may form a bound state (15) of $B_{1g}(d_{x^2-y^2})$ symmetry (46). Because of phase fluctuations of the SC order parameter, global superconductivity may occur at $T_c$ well below the binding temperature (26, 27). Geshkenbein et al. discussed superconductivity resulting from the Bose condensation of preformed pairs with no dispersion in the vicinity of $\{k_{an}\}$ points coexisting and weakly interacting with unpaired fermions on other patches of the Fermi surface (47).

The 75-meV mode observed in Raman spectra puts additional constraints on the theoretical models: The 75-meV excitation is related to the pseudogap phenomena, and it has $B_{1g}(d_{x^2-y^2})$ symmetry, a long lifetime, a few percent of the 2-M scattering intensity, and intensity enhancement at $T_c$. Moreover, the relation to the QP energies seen in one-electron spectroscopies suggests that the chemical potential of the system is in the middle of the QP (pseudo)gap.

All of the properties observed by Raman, ARPES, and tunneling spectroscopies reveal an increase in the coherence of the electronic system as the underdoped cuprates are cooled toward $T_c$. The rearrangement of the Raman spectra with formation of the 75-meV mode is an indication that a partially coherent state forms above $T_c$ out of the incoherent QPs in the vicinity of $\{k_{an}\}$ points. This long-lived state might consist of bound states of doped holes (preformed Cooper pairs) or of more complex many-body objects. Light scattering may break up the bound state into unbound QPs by a process similar to 2-M scattering. We propose that this process, enhanced by a final-state resonance, is the origin of the 600 cm$^{-1}$ $B_{1g}$ Raman peak, and we suggest that it should be present for underdoped materials with AF correlations sufficient to exhibit underdamped 2-M excitations. The AF correlations that cover a few lattice spacings should last for at least 10$^{-13}$ s, the lifetime of the bound state. The ratio of 2-M scattering to the 600 cm$^{-1}$ mode intensity may represent a rough estimate of the number of sites with AF ordered Cu spins relative to the number in the bound state.

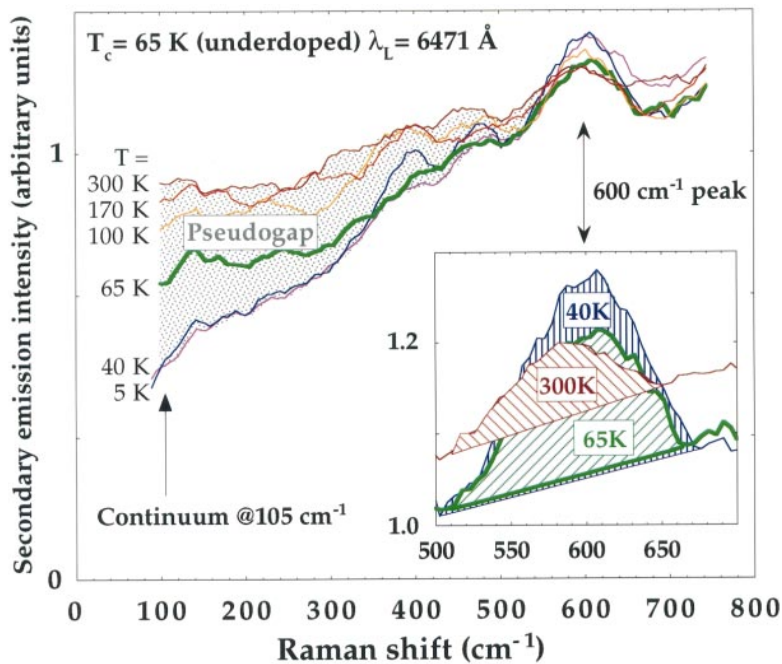

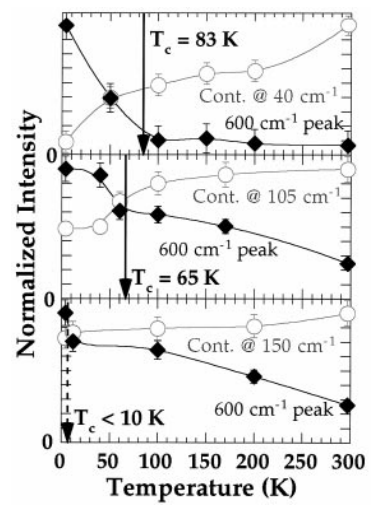

**Fig. 3.** The $B_{1g}$ Raman scattering spectra for underdoped $T_c$ = 65 K Bi$_2$Sr$_2$CaCu$_2$O$_{8\pm\delta}$ as a function of temperature. The temperature dependence of the area under 600 cm$^{-1}$ mode above the continuum is shown in the inset.

**Fig. 4.** Temperature dependence of the normalized low-frequency continuum intensity (circles) and a normalized (integrated above the continuum) 600 cm$^{-1}$ peak intensity (diamonds) for the $T_c$ < 10 K, $T_c$ = 65 K, and $T_c$ = 83 K underdoped samples. Vertical arrows denote $T_c$.

Within the bound-state scenario, the essential point is: The bound state is compatible with the surrounding AF order (13, 15, 26, 46). This makes formation of the long-lived 600 cm$^{-1}$ mode possible. Occupancy of the collective bound state lowers the density of the single holes, leading to reduction of the low-frequency inverse lifetime, suppression of the incoherent low-frequency DOS, and formation of the pseudogap. Because of different couplings to light and also because of the very different lifetimes of single holes and the bound state, there is no conservation of spectral weight for Raman scattering: The low-frequency intensity reduction is stronger than the increase of the intensity of the 600 cm$^{-1}$ mode. As seen from Figs. 2, A to C, and 3, the low-temperature reduction of the scattering rate, as also shown by optical data (8), occurs only below the 600 cm$^{-1}$ binding energy. Cooling increases the concentration in the bound state and enhances the pseudogap. At a critical density, the bound state manifests phasing and condenses into the collective SC state. Simultaneously, because of the phasing effect, the Raman spectra acquire additional 600 cm$^{-1}$ peak intensity, and therefore the peak enhancement reflects the fraction of the bound carriers in the SC condensate.

For the very underdoped sample ($T_c < 10$ K), only a weak pseudogap develops, and the integrated 600 cm$^{-1}$ peak intensity exhibits a weak enhancement (Fig. 4). Both the pseudogap and the peak enhancement are stronger for the $T_c = 65$ K sample. The sample with slightly higher doping ($T_c = 83$ K) exhibits an effective condensation into the SC state, with formation of a 2$\Delta$ peak–like feature in the Raman spectra out of the 600 cm$^{-1}$ peak. Simultaneously with a strong increase in the density of the coherent bound state, a well-defined QP peak develops in the ARPES and tunneling spectra at about half of the 2$\Delta$-peak energy [see figure 1 in (9) and figure 1 in (45)]. However, we believe that for samples with lower $T_c$ or for the $T_c = 83$ K sample at temperatures above $T_c$, where both the low-frequency reduction of incoherent intensity and the QP peak enhancement caused by phasing are not very pronounced, the lower energy resolution of ARPES spectroscopy prevents the observation of the weak QP peak on the strong incoherent background. Also, for all underdoped samples, only a partial condensation of single holes occurs even at the lowest temperature. Conventional Fermi liquid-based Raman scattering theory for $d$-wave superconductors predicts a cubic power law for the low-frequency tail of the $B_{1g}$ scattering intensity (37). For the $T_c = 83$ K sample, which shows a pronounced SC condensation, an approximate cubic power law is observed for frequencies only below the shoulder at ~200 cm$^{-1}$. The shoulder may result from a SC gap opening along small pocket-type Fermi surfaces observed by ARPES around ($\pm\pi/2a$, $\pm\pi/2a$) (11). A weak 2$\Delta$ peak–like feature has been observed at ~200 cm$^{-1}$ for $B_{2g}$ scattering geometry that is most sensitive for excitations along the pocket-type Fermi surfaces (48).

In contrast, overdoped materials show a different picture. Even though normal state scattering is still mostly incoherent, the scattering intensity above $T_c$ is weaker than for underdoped materials (compare Figs. 2B and 2D). There is no robust excitation near 600 cm$^{-1}$. A strong coherent 2$\Delta$ peak, much stronger than the 600 cm$^{-1}$ peak for the underdoped materials, develops below $T_c$. The peak shape, including its low-frequency tail, may be reproduced by Fermi liquid-based Raman scattering theory for $d$-wave SC with a large Fermi surface (37). This result demonstrates that (i) in agreement with ARPES studies (11), a large Fermi surface is restored for materials above optimal doping, and (ii) at least in the SC state, the incoherent scattering is greatly reduced up to 400 cm$^{-1}$.

# Nearly Singular Magnetic Fluctuations in the Normal State of a High-$T_c$ Cuprate Superconductor

G. Aeppli, T. E. Mason,* S. M. Hayden, H. A. Mook, J. Kulda

Polarized and unpolarized neutron scattering was used to measure the wave vector– and frequency-dependent magnetic fluctuations in the normal state (from the superconducting transition temperature, $T_c$ = 35 kelvin, up to 350 kelvin) of single crystals of $La_{1.86}Sr_{0.14}CuO_4$. The peaks that dominate the fluctuations have amplitudes that decrease as $T^{-2}$ and widths that increase in proportion to the thermal energy, $k_BT$ (where $k_B$ is Boltzmann's constant), and energy transfer added in quadrature. The nearly singular fluctuations are consistent with a nearby quantum critical point.

The normal state of the metallic cuprates is as unusual as their superconductivity. For example, the electrical resistivity of samples with optimal superconducting properties is linear in temperature ($T$) from above 1000 K to the superconducting transition temperature, $T_c$ (*1*). Correspondingly, infrared reflectivity reveals charge fluctuations with a characteristic energy scale that is proportional only to $T$ (*1, 2*). Furthermore, the effective number of charge carriers, as measured with the classic Hall effect, is strangely $T$-dependent. Even so, the Hall angle, a measure of the deflection of carriers in the material by an external magnetic field, follows a $T^{-2}$ law (*3*). Thus, the metallic charge carriers in the doped cuprates exhibit peculiar but actually quite simple properties (*4*) in the normal state. Moreover, these properties do not vary much between the different high-$T_c$ families.

Electrons carry spin as well as charge, so it is reasonable to ask whether the normal state magnetic properties derived from the spins are as simple and universal as those derived from the charges. Experiments to probe the spins include classical magnetic susceptometry, where the magnetization in response to a homogeneous external magnetic field is measured, and resonance experiments, where nuclear dipole and quadrupolar relaxation is used to monitor the atomic-scale magnetic fluctuations. The spin-sensitive measurements yield more complex and less universal results than those sensitive to charge, and do not seem obviously related to the frequency-dependent conductivity $\sigma(\omega,T)$ (where $\omega$ is frequency), probed in electrical, microwave, and optical experiments. In particular, there is little evidence for magnetic behavior that is as nearly singular in the sense of diverging (for $T \to 0$) amplitudes, time constants, or length scales, as the behavior of $\sigma(\omega,T)$.

We report here nearly singular behavior of the magnetic fluctuations in the simplest of high-$T_c$ materials, namely, the compound $La_{2-x}Sr_xCuO_4$, whose fundamental building blocks are single $CuO_2$ layers. The experimental tool was inelastic magnetic neutron scattering. A beam of mono-energetic neutrons is first prepared and then scattered from the sample, and the outgoing neutrons are labeled according to their energies and directions to establish an angle and energy-dependent scattering probability. Because the neutron spin and the electron spins in the sample interact through magnetic dipole coupling, the cross section is directly proportional to the magnetic structure function, $S(\mathbf{Q},\omega)$, the Fourier transform of the space- and time-dependent two-spin correlation function. The momentum and energy transfers $\mathbf{Q}$ and $\omega$ are simply the differences between the momenta and the energies of the ingoing and outgoing neutrons, respectively. According to the fluctuation-dissipation theorem, $S(\mathbf{Q},\omega)$ is in turn proportional to the imaginary part, $\chi''(\mathbf{Q},\omega)$, of the generalized linear magnetic response $\chi(\mathbf{Q},\omega)$. The bulk susceptibility measured with a magnetometer is the long-wavelength, small-wavenumber, ($\mathbf{Q}\to 0$), limit of $\chi'(\mathbf{Q},\omega = 0)$, and the nuclear resonance techniques yield averages of $\chi''(\mathbf{Q},\omega \sim 0)$ over momenta $\mathbf{Q}$, which are of order inverse interatomic spacings.

Figure 1A is a schematic phase diagram for $La_{2-x}Sr_xCuO_4$ as a function of $T$, hole doping ($x$), and pressure ($y$). Holes and pressure are generally introduced chemically, most notably through substitution of $Sr^{2+}$ and $Nd^{3+}$ ions, respectively, for the $La^{3+}$ ions in $La_2CuO_4$ (*5, 6*). Possible magnetic ground states range from simple antiferromagnetic (AFM for $x \sim 0$) to a long-period spin density wave with strong coupling to the underlying lattice (shown as a gray "mountain" for $x \sim 0.1$ in Fig. 1A). Unit cell doubling, where the spin on each $Cu^{2+}$ is antiparallel to those on its nearest neighbors displaced by vectors $(0, \pm a_o)$ and $(\pm a_o, 0)$ in the (nearly) square $CuO_2$ planes, characterizes the simple AFM state (*7*); the lattice constant, $a_o$ = 3.8 Å. The associated magnetic Bragg peaks, observed by neutron scattering, occur at reciprocal lattice vectors $\mathbf{Q}$ of the form $(n\pi, m\pi)$, where $n$ and $m$ are both odd integers; the axes of the reciprocal lattice coordinate system are parallel to those of the underlying square lattice in real space.

Substitution of $Sr^{2+}$ for $La^{3+}$ introduces holes into the $CuO_2$ planes and initially replaces the AFM phase by a magnetic (spin) glass phase. It is in this nonsuperconducting composition regime, for which the magnetic signals are strong and large single crystals have long been available, that the most detailed $T$-dependent magnetic neutron scattering studies have been performed (*8*). With further increases in $Sr^{2+}$ content, the magnetic glass phase disappears and superconductivity emerges. At the same time, the commensurate peak derived from the order and fluctuations in the nonsuperconducting sample splits into four incommensurate peaks, as indicated in Fig. 1B (*9*). These peaks are characterized by a position, an amplitude, and a width. Earlier work (*9*) has described how the peak positions vary with composition at low temperatures. Our contribution is to follow the red trajectory in Fig. 1 and thus obtain the $T$ and $\omega$ dependence of the amplitude and width, which represent the maximum magnetic response and inverse magnetic coherence length, respectively.

The $La_{1.86}Sr_{0.14}CuO_4$ crystals used here are the same as those used in our determina-

G. Aeppli, NEC Research Institute, 4 Independence Way, Princeton, NJ 08540, USA, and Risø National Laboratory, 4000 Roskilde, Denmark.
T. E. Mason, Department of Physics, University of Toronto, Toronto, Canada, M5S 1A7 and Risø National Laboratory, 4000 Roskilde, Denmark.
S. M. Hayden, Department of Physics, University of Bristol, Bristol BS8 1TL, UK.
H. A. Mook, Oak Ridge National Laboratory, Oak Ridge, TN 37831, USA.
J. Kulda, Institut Laue-Langevin, BP 156X, Grenoble Cedex, France.

*To whom correspondence should be addressed. E-mail: thom.mason@utoronto.ca